\documentclass[aps,prl,twocolumn,superscriptaddress,amsmath,amssymb]{revtex4-1}
\usepackage{color}
\usepackage{graphicx}
\usepackage{cancel}
\usepackage{amsfonts}
\usepackage{amssymb}
\usepackage{amsthm}
\usepackage{ulem}
\usepackage{url}
\usepackage{hyperref}
\usepackage{float}
\usepackage{multirow}  
\usepackage{dcolumn}   
\usepackage{bm}        
\usepackage{enumerate}
\usepackage{mathtools}

\usepackage[dvipsnames]{xcolor}
\hypersetup{
	colorlinks,
	linkcolor={green!50!black},
	urlcolor={blue!80!black},
	citecolor   = {red!50!black} 
}

\def  \bsig    {\mbox{\boldmath$\sigma$}}
\def  \bL    {\mbox{\boldmath$\Lambda$}}

\def \bk       {\mbox{\boldmath\footnotesize$\mathrm{k}$}}
\def \bq       {\mbox{\boldmath\footnotesize$\mathrm{q}$}}

\begin{document}

\title{Theory of bi-linear magnetoresistance within the minimal model for surface states in topological insulators
}

\author{A. Dyrda\l}
\affiliation{Faculty of Physics, Adam Mickiewicz University, 61-614 Pozna\'n, Poland}
\affiliation{Institut für Physik, Martin-Luther-Universität Halle–Wittenberg, 06099 Halle (Saale), Germany}
\author{J. Barna\'s}
\affiliation{Faculty of Physics, Adam Mickiewicz University, 61-614 Pozna\'n, Poland}
\author{A. Fert}
\affiliation{Unit\'{e} Mixte de Physique, CNRS, Thales, Univ. Paris-Sud, Universit\'{e} Paris-Saclay, 91767, Palaiseau, France}

\begin{abstract}
A new mechanism of bi-linear magnetoresistance (BMR)  is studied theoretically within the minimal model describing  surface electronic states in topological insulators (TIs). The BMR  appears as a  consequence of the second-order response to electric field, and depends linearly on both electric field (current) and magnetic field. The mechanism is  based on the interplay of current-induced spin polarization and scattering processes due to  peculiar spin-orbit defects. The proposed mechanism is compared to that based on a Fermi surface warping, and is shown to be dominant at lower Fermi  energies. We provide a consistent theoretical approach based on the Green function formalism and show that the magnetic field dependent relaxation processes in the presence of non-equilibrium current-induced spin polarization give rise  to the BMR.
\end{abstract}

\date{\today}
\maketitle

\textit{Introduction.} -- Simultaneous breaking of space and time inversion symmetry  may lead to a variety of novel phenomena associated with  nonreciprocal system response to external driving forces.
In such a case (e.g. in noncentrosymmetric crystals in external magnetic field $\mathbf{b}$) the resistance may possess not
only the $b$-squared  but also the
bi-linear term in the resistivity~\cite{Rikken2001,Rikken2005,Nagaosa2018}, which
 indicates that reversal of the external magnetic field  is equivalent to the reversal of the current flow direction, i.e.
$R(\mathbf{j}, \mathbf{b}) =R(-\mathbf{j}, -\mathbf{b}) \ne R(\mathbf{j}, -\mathbf{b}), R(-\mathbf{j}, \mathbf{b})$. The non-reciprocal
bi-linear term in the magnetoresistance attracts recently much attention, being not only intriguing from the fundamental physics point of view,
but also due to possible applications in new-generation spintronics devices based on  spin-to-charge interconversion phenomena.

The discovery of giant and tunneling magnetoresistance~\cite{Baibich1988,Binach1989,Julliere}
initiated new routes in the development of data storage technologies and spintronics devices.  Due to the continued tendency towards  miniaturization of nanoelectronic elements, there is  a need to find an effective way of electron spin control in ultra-thin layers or in two-dimensional crystals.  Therefore, the main attention is currently focused on the spin-orbit driven phenomena that provide efficient spin-to-charge interconversion, even in non-magnetic materials  \cite{Soumyanarayanan2016}. The most prominent examples are the spin Hall effect~\cite{Dyakonov1971,Hirsch1999,engel07,sinova15} and the current-induced spin polarization (CISP)
~\cite{aronov89,Edelstein,Ganichev2001,Ganichev2002}.
The spin-orbit interaction also plays a crucial role in the anisotropic magnetoresistance (AMR)~\cite{McGuire1975}. By analogy between the spin Hall and anomalous Hall effects (the latter being a  counterpart of the SHE in magnetic systems), an essential research topic in spintronics is a search for the counterpart of AMR phenomenon in nonmagnetic materials.

\begin{figure}[t]
	\includegraphics[width=0.8\columnwidth]{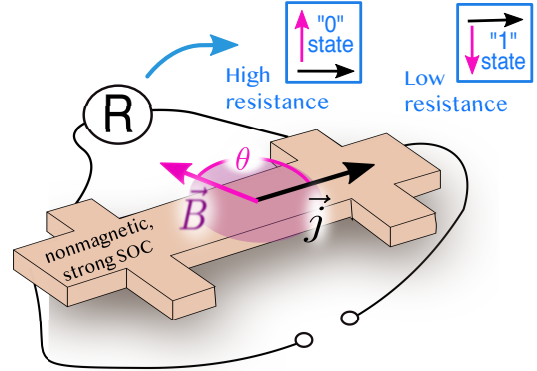}
	\caption{ Schematic picture of the system under consideration. The angle $\theta$ is defined as the angle between the orientation of  charge current density $\mathbf{j}$ and external magnetic field $\mathbf{b}$. Two states corresponding to antiparallel orientations of the magnetic field (but normal to current) have different resistances and may play the role as two logical states "0" and "1".
	}
	\label{fig:Pict}
\end{figure}

In recent  years, new concepts for the magnetoresistance phenomena mediated by spin-orbit coupling have been reported. One of them relies on connecting ferromagnet (insulating or metallic) with a material which exhibits strong spin-orbit coupling (e.g., heavy metals) and explore the interfacial processes induced by the spin Hall effect. Such a proximity-induced magnetoresistance is referred to as the spin Hall magnetoresistance  (SMR)~\cite{NakayamaSaitoh2013,NakayamaBauer2013,KimHayashi}. In a variety of hybrid structures, where the proximity-induced magnetism affects the electronic states of the material with strong spin-orbit coupling, or in magnetic topological insulators(TIs), the magnetoresistance reveals a unique unidirectional sensitivity to the relative orientation of the magnetization and charge current~\cite{Avci_nature2015,olejnik2015,Avci_APL2015,Tokura2016,Lv,Vignale2016}.

The bilinear magnetoresistance (BMR) has been measured recently in nonmagnetic systems, e.g., in three-dimensional topological insulators~\cite{He2018}  and two-dimensional electron gas (2DEG) formed at interfaces of perovskite oxides~\cite{Narayanapillai2017}.
The unidirectional nature of magnetoresistance  in these systems is undoubtedly related to the nonlinear response to electric field. Recently Zhang and Vignale~\cite{He2018,Vignale2018} have provided a theoretical description of this phenomenon, according to which the BMR is the consequence of hexagonal warping of the band structure, that leads to a de-freezing of the spin-momentum locking, enabling thus electron backscattering.   However, recent experimental data on the topological insulator $\alpha$-Sn(001)~\cite{Rojas} have shown~\cite{PC} that BMR can exist in materials without hexagonal symmetry and thus cannot be explained by the hexagonal warping. Therefore one may expect there is another  mechanism contributing to
the BMR phenomenon.

In this letter we formulate such a mechanism and present the relevant theoretical description
which is  based on a
minimal model of TIs, i.e. the model which does not include the hexagonal warping term.  
Instead of this, we take into account scattering by spin-orbit inhomogeneities  and the related magnetic field dependence of the scattering rate.
We show that this leads to BMR, and
this contribution is dominant at lower Fermi energies.

\vspace{0.2cm}
\textit{Surface states of TIs under external fields.} -- Assuming the minimal model describing surface states in TIs (thus neglecting the hexagonal warping and $\mathbf{k}$-quadratic terms) one can write Hamiltonian of the system in external electric and magnetic fields in the $\mathbf{k}$ basis as
\begin{equation}
\label{Htot}
\hat{H}^{\scriptstyle{tot}}_{\mathbf{k} \mathbf{k}'} =
(\hat{H}^{0}_{\mathbf{k}} + \hat{H}^{\mathbf{A}}_{\mathbf{k}})\delta_{\mathbf{k} \mathbf{k}'}  + \hat{V}^{\mathrm{sc}}_{\mathbf{k} \mathbf{k}'},
\end{equation}
where
$\hat{H}^{\mathbf{A}}_{\mathbf{k}}$
describes interaction of the system with dynamical electric field, $\hat{H}^{\mathbf{A}}_{\mathbf{k}} = - e \hat{\mathbf{v}}_{\mathbf{k}}\cdot \mathbf{A}$
(here  $e$ denotes the electron charge, $\mathbf{A}$ is the electromagnetic vector potential, and $\hat{\textbf{v}}$ is the velocity operator defined as $\hat{\textbf{v}}_{\mathbf{k}} = \hbar^{-1} \nabla_{\mathbf{k}} \hat{H}^{0}_{\mathbf{k}}$),  $\hat{V}^{\mathrm{sc}}_{\mathbf{k} \mathbf{k}'}$
accounts for scattering on structural defects (to be specified later), and $\hat{H}^{0}_{\mathbf{k}}$ has the form
\begin{equation}
\label{H0}
\hat{H}^{0}_{\mathbf{k}}  =  v (\mathbf{k}\times \hat{\mathbf{z}})\cdot \bsig + \mathbf{B} \cdot \bsig + \mathcal{J} \mathbf{S}\cdot \bsig .
\end{equation}
The first term in $\hat{H}^{0}_{\mathbf{k}}$
describes pure surface states of TIs  and has the well known form~\cite{Adam,ozturk}
with $v = \hbar v_{F}$ ($v_{F}$ is the Fermi velocity), $\mathbf{k}$  being  the wave vector, $\hat{\mathbf{z}}$ standing for the unit vector normal to the surface, and $\bsig = (\sigma_{x}, \sigma_{y}, \sigma_{z})$ denoting the vector of Pauli matrices acting in the spin space.
The second term of $\hat{H}^{0}_{\mathbf{k}}$ describes the influence of an external in-plane magnetic field
 $\mathbf{B} = (B_{x}, B_{y}, 0)$ (the field is measured in the energy units).
The last term in $\hat{H}^{0}_{\mathbf{k}}$ takes into account effective coupling between the electrons and externally-induced spin polarization $\textbf{S}$. Such a spin polarization is driven by an external electric field and
appears only in the presence of spin-orbit  coupling in the system. This term is written in the Heisenberg-like form,
where the coupling constant $\mathcal{J}$ for the surface states of TIs may be expressed by the formula $\mathcal{J} =  -8 \pi v_{F}/ k_{F}$  for the assumed spin chirality (see the  Supplementary Material \hyperref[S1]{ S1\ref{S1}}).
Since $\mathbf{S}$ is a linear function of $\mathbf{j}$,  one can see immediately that this term
introduces the  uni-directional character (with respect to $\mathbf{j}$) of the system response.
It is also worth to note that the non-equilibrium spin polarization $\mathbf{S}$ may interact with local magnetic moments, leading to a  spin-orbit torque (see, e.g. Refs. [\onlinecite{Manchon2008,Manchon2009,MatosAbiague,Miron2011,Kurebayashi2014}]).
Hereinafter, without loosing generality of our analysis, we assume that external electric field is oriented in the $x$-direction, $\mathbf{E} = (E, 0, 0)$,  as indicated in Fig.1. Accordingly, the last term in Eq.~(\ref{H0})
can be written as $\mathcal{J} S\sigma_{y}$ (see \hyperref[S1]{ S1\ref{S1}}) while $H^{\mathbf{A}}_{\mathbf{k}}$  takes the form  $H^{\mathbf{A}}_{\mathbf{k}} = - e v_{\mathbf{k}x} A_{x}$.

The last two terms of Hamiltonian (2) can be written as $\mathbf{B}_{\rm eff}\cdot\bsig$, with  $\mathbf{B}_{\rm eff}= \mathbf{B} + \mathcal J \mathbf{S}$. The effective field $\mathbf{B}_{\rm eff}$ shifts the Dirac cone out of the Brillouin zone center.  However,
the static effective in-plane field $\mathbf{B}_{\rm eff}$
can be removed from the Hamiltonian~(\ref{H0}) by the gauge transformation~\cite{Bauer2017}: $\mathbf{k} \to \mathbf{q} - \frac{e}{\hbar} \bL $, where $\bL= \frac{\hbar}{v e} \mathbf{B}_{\rm eff}\times \hat{\mathbf{z}}$.
Upon the  transformation, Hamiltonian of the system takes the following form:
\begin{eqnarray}
\hat{H}^{tot}_{\mathbf{q} \mathbf{q}'} = (\hat{H}^{0}_{\mathbf{q}} + \hat{H}^{\mathbf{A}}_{\mathbf{q}})\delta_{\mathbf{q},\mathbf{q}'}  +\hat{V}_{\mathbf{q} \mathbf{q}'}^{\mathrm{sc}}   ,\\
\hat{H}^{0}_{\mathbf{q}} = v(\mathbf{q} \times \hat{\mathbf{z}})\cdot \bsig .\hspace{0.5cm}
\end{eqnarray}
Although the effective magnetic field has been removed from the Hamiltonian $\hat{H}_{0}$, it still appears in the scattering part $\hat{V}_{\mathbf{q} \mathbf{q}'}^{\mathrm{sc}}$ for some class of structural defects, as will be shown in the following.
 We emphasize, that we consider conduction by a single surface without any contribution from the opposite surface.

\vspace{0.2cm}
\textit{Relaxation time and conductivity.} -- We begin with relaxation processes and their dependence on the magnetic and electric fields. Such a dependence is crucial for BMR.  Since scattering by defects with pure electrostatic potential does not depend on magnetic field (also upon the gauge transformation), we consider structural defects that include spin-orbit interaction. As we will see below, scattering by such defects depends on magnetic field upon the gauge transformation.
Assuming local deviation of the parameter $v$ in the Hamiltonian~(\ref{H0}) due to local random Rashba-like spin-orbit field, one may write the scattering term in the form \cite{Winkler,Sherman,Golub,Japaridze,Braatas}:
\begin{equation}
\hat{V}^{\rm sc}(\mathbf{r}) = - \frac{i}{2} \left\{ \nabla_{y}, \alpha(\mathbf{r})\right\} \sigma_{x} +  \frac{i}{2} \left\{ \nabla_{x}, \alpha(\mathbf{r})\right\} \sigma_{y}.
\end{equation}
The parameter $\alpha(\mathbf{r})$ describes local deviations of the spin-orbit coupling due to local defects distributed randomly in the structure. For the  white noise distribution of the defects one may write:
$\langle \alpha(\mathbf{r})\rangle = 0$,  $\langle \alpha(\mathbf{r}) \alpha(\mathbf{r}')\rangle = n_{i} \alpha^{2} \delta(\mathbf{r} - \mathbf{r}')$, and $\langle |\alpha_{\mathbf{k} \mathbf{k}'} |^{2} \rangle= n_{i} \alpha^{2}$, where $n_i$ is the concentration of  scattering centers and $\alpha$ is treated as a parameter.  Thus, the scattering  potential in the momentum space takes the form
\begin{equation}
\label{H_alpha}
\hat{V}^{\rm sc}_{\mathbf{k} \mathbf{k}'}  = \frac{n_i\alpha^2}{2} \left[ (k_{y} + k_{y}') \sigma_{x} - (k_{x} + k_{x}') \sigma_{y} \right].
\end{equation}
Upon the gauge transformation introduced above, the scattering potential  (\ref{H_alpha}) may be written as
\begin{eqnarray}
\label{Halpha_transf}
 \hat{V}^{\rm sc}_{\mathbf{q} \mathbf{q}'} = \frac{n_i\alpha^2}{2} \left[ (q_{y} + q_{y}') \sigma_{x} - (q_{x} + q_{x}') \sigma_{y}\right]\nonumber\\ - \frac{n_i\alpha^2}{v} \left[ B_{x} \sigma_{x} + (B_{y}+ \mathcal{J} S_{y}) \sigma_{y} \right].
\end{eqnarray}

 To determine transport properties
 we use the Green's function formalism and calculate first the self-energy which allows to find both the relaxation time and  averaged Green function. In the  Born approximation the self-energy is defined by the expression
\cite{mahan,agd}:
\begin{eqnarray}
\label{SE-def}
\Sigma_{\mathbf{q}}^{R}(\varepsilon)
=   \int \frac{d^{2} \mathbf{q}'}{(2\pi)^{2}}  \hat{V}^{\mathrm{sc}}_{\mathbf{q} \mathbf{q}'} G_{\mathbf{q}'}^{0R} \, \hat{V}^{\mathrm{sc}}_{\mathbf{q}' \mathbf{q}},
\end{eqnarray}
where $ G_{\bq}^{0R} $ denotes the retarded Green function of the Hamiltonian $\hat{H}_{0}$,  ${G_{\bq}^{0R} = [(\varepsilon+ i \eta) \sigma_{0} - \hat{H}^{0}_{\mathbf{q}}]^{-1}}$. In the weak scattering limit the self-energy can be calculated analytically. For scattering due to random Rashba-like potential (see supplementary notes \hyperref[S2]{S2\ref{S2}}),
the relaxation rate $\Gamma$ has the following form:
\begin{eqnarray}
\Gamma_{\bq} = \frac{\gamma_{0}}{4 v^{2}} \bigglb[ (|\varepsilon| + v q)^{2} + 8 B_{y} \mathcal{J} S_{y} + 4 B^{2} \biggrb. \hspace{1cm}\nonumber\\
+ \bigglb.4 (|\varepsilon| + q v) \left( (B_{y} + \mathcal{J} S_{y}) \frac{q_{x}}{q} - B_{x} \frac{q_{y}}{q} \right) \biggrb].
\end{eqnarray}
At the mass surface, $v q =  |\varepsilon| $, this expression takes the form:
\begin{eqnarray}
\Gamma(\phi)|_{q = \frac{|\varepsilon|}{v}}  = \Gamma_{0} \left[ 1 + 2 \frac{B_{y} \mathcal{J} S_{y}}{\varepsilon^{2}} + \frac{B^{2}}{\varepsilon^{2}} \right.\hspace{1cm}\nonumber\\
+\left. \frac{2}{|\varepsilon|} \left[ (B_{y} + \mathcal{J} S_{y}) \cos\phi - B_{x} \sin\phi\right]\right],
\end{eqnarray}
where $\Gamma_{0} = n_{i} \alpha^{2} \frac{|\varepsilon|^{3}}{4 v^{4}}$. Note the relaxation rate depends on the angle $\phi$ between the wavevector ${\bq}$ and the axis $x$. Since the relaxation time is given by the relation $\tau = \hbar/2\Gamma$,  and the external magnetic field is small in comparison to the spin-orbit field in TIs, one can write:
\begin{eqnarray}
\tau(\phi) = \frac{\hbar}{2 \Gamma_{0}} \left[ 1 - 2 \frac{B_{y} \mathcal{J} S_{y}}{\varepsilon^{2}} - \frac{B^{2}}{\varepsilon^{2}} \right.\hspace{1.5cm}\nonumber\\
-\left. \frac{2}{|\varepsilon|} \left[ (B_{y} + \mathcal{J} S_{y}) \cos\phi - B_{x} \sin\phi\right]\right].
\end{eqnarray}

Taking into account that $S \sim j \sim E$, the relaxation time depends not only on the strength of magnetic field $B$ but also on the electric field (or  electric current density) through the nonequilibrium spin density $ S_{y}=\frac{\hbar^2}{2ev}j_x$, and on the angle $\theta$ between the axis $x$ and in-plane magnetic field (defined by the relation  $B_y=B \sin \theta$). This is an important result which  shows that the semiclassical treatment based on a constant relaxation time approximation may not be sufficient for description of the magnetoelectric phenomena. This is not only in the case of systems with anisotropic energy dispersion~\cite{Vyborny,Trushin}  (anisotropic Fermi contours), but also for systems with isotropic energy dispersion when the space-inversion and time-inversion symmetries are broken.

The electrical dc conductivity  in the linear response to
$\hat{H}^{A}$  can be determined from the formula
\begin{equation}
\label{sigma_xx}
\sigma_{xx} = \frac{e^{2} \hbar}{2\pi} \int \frac{d^{2} \mathbf{q}}{(2\pi)^{2}} \mathrm{Tr}\left\{\hat{v}_{\bq x} \bar{G}_{\bq}^{R}(\varepsilon_{F}) \mathcal{V}_{\bq x} \bar{G}_{\bq}^{A}(\varepsilon_{F})\right\} .
\end{equation}
Here $\bar{G}_{\bq}^{R/A}(\varepsilon) = [\varepsilon \sigma_{0} - \hat{H}^{0}_{\bq} - \Sigma_{\bq}^{R/A}]^{-1}$ is the disorder averaged retarded/advanced Green function at the Fermi level ($\varepsilon = \varepsilon_{F}$), and $\mathcal{V}_{\bq x}$ is the velocity vertex function in the ladder approximation (see
 the supplementary material \hyperref[S3]{S3\ref{S3}} for details). The  final result for the longitudinal conductivity acquires the form:
\begin{eqnarray}
\sigma_{xx} = \sigma_{xx}^{0} \left[1 - \frac{15}{2} \frac{B_{y} \mathcal{J} S_{y}}{\varepsilon_{F}^{2}} - 3 \frac{B^{2}}{\varepsilon_{F}^{2}}  \right]\hspace{0.7cm}\nonumber \\
 =  \sigma_{xx}^{0}\left[1 + 15 \frac{h}{e k_{F}} \frac{j_x B}{\varepsilon_{F}^{2}} \sin\theta - 3 \frac{B^{2}}{\varepsilon_{F}^{2}}  \right],
\end{eqnarray}
where $\sigma^{0}_{xx} = \frac{e^{2}}{h} \frac{|\varepsilon_{F}|}{8 \Gamma_{0}}$ is the longitudinal conductivity in the absence of external magnetic field, while $j_x$ is the density of current flowing parallel ($j_x=j$ ) or antiparallel ($j_x=-j$) to the $x$-axis.  In addition, the explicit expression for the coupling constant $\mathcal{J}$ has been taken into account.

\vspace{0.2cm}
{\textit{Discussion.}} --
To describe magnetoresistance (MR) we use the  conventional definition, $MR = [\rho (B) - \rho (B=0)]/\rho (B=0) = [\sigma (B=0) / \sigma (B)]- 1$. Then,
one can introduce two other characteristics describing separately the  symmetric and antysymmetric parts of MR. The symmetric part may be written  as
${sMR = \left[ MR(j_x=j) + MR(j_x=-j)\right]/2}$. In turn, the antisymmetric MR contains the information about the unidirectional character of transport in the system and is determined by the bilinear term in conductivity/resistivity. One can define this bilinear magnetoresistance as  ${BMR =\left[ MR(j_x=j) - MR(j_x=-j)\right]/2}$.

From the formula for longitudinal conductivity one finds explicit expressions for the BMR and sMR. To make easier  comparison with the corresponding formulas available in the literature, we write explicitly $\mathbf{B}=g\mu_B\mathbf{b}$, where $\mathbf{b}$ is the magnetic field. As a result one finds:
\begin{eqnarray}
BMR = \mathcal{A}_{\rm BMR}\frac{j_x}{j}\sin{\theta} \equiv
15 \frac{h}{|e|} \frac{\mu_{B} g}{k_{F} \varepsilon_{F}^{2}} j b\frac{j_x}{j}\sin {\theta} \nonumber \\
= \left(30 \pi g \mu_{B} \frac{\hbar^{2}}{|e|}\right)
 \frac{v_{F}}{|\varepsilon_{F}|^{3}} j b\frac{j_x}{j} \sin {\theta},
\end{eqnarray}
for the bilinear magnetoresistance, where $\mathcal{A}_{\rm BMR}$ is the amplitude of BMR, and
\begin{equation}
sMR =
(3 g^{2} \mu_{B}^{2}) \frac{b^{2}}{\varepsilon_{F}^{2}}
\end{equation}
for the symmetric magnetoresistance. The BMR varies with the magnetic field orientation as $\sin \theta$. This is shown in Fig.2a and Fig.2b for several values of the field and for $j_x=j$ (a) and $j_x=-j$ (b). The amplitude of BMR grows linearly with $b$ and also with $j$, see Figs.2(c,d) for indicated parameters.  In turn, the symmetric MR  is independent of the field orientation (within the model assumed in the description), and depends on the strength of magnetic field as $b^2$.
Therefore it is also referred to as quadratic magnetoresistance (QMR), see also Fig.2(f). Since QMR varies with magnetic field as $b^2$ while BMR grows linearly with $b$, the ratio of QMR and amplitude of BMR is a linear function of $b$. Moreover, for typical values of $j$ and $b$, the QMR is remarkably smaller than BMR.
Thus, based on magnetoresistance measurements one can  estimate
parameters describing the electronic band structure of TIs, such as $k_{F}$, $v$, or $g$.

\begin{figure}[t]
	\includegraphics[width=0.9\columnwidth]{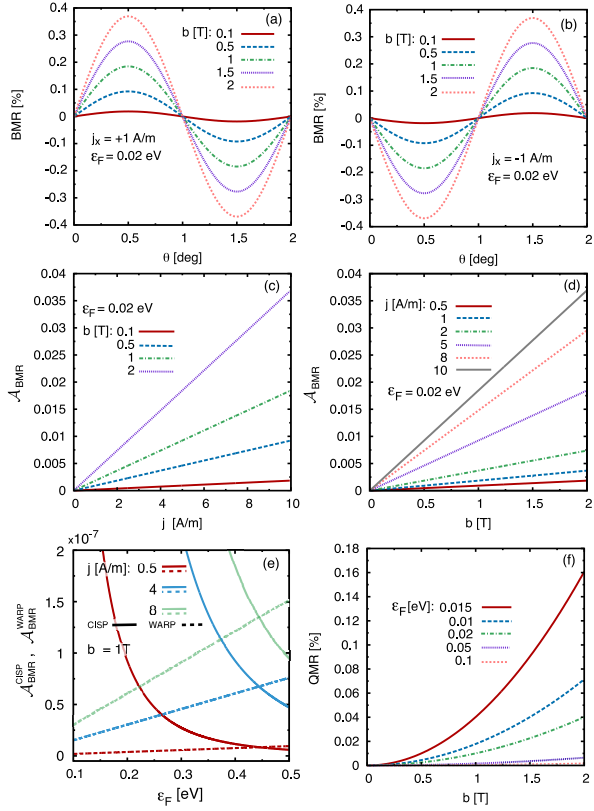}
	\caption{ BMR as a function $\theta$ for indicated values of magnetic field and for $j_x>0$ (a) and $j_x<0$ (b); amplitude $\mathcal{A}_{BMR}$ of BMR as a function of current $j$ (c) and as a function of magnetic field $b$ (d); comparison of the BMR amplitude in the model based on warping, $\mathcal{A}_{BMR}^{\rm warp}$, with the amplitude based on CISP, $\mathcal{A}_{BMR}^{\rm CISP}$ (e); quadratic magnetoresistance  as a function of magnetic field (f). }
	\label{fig:Pict}
\end{figure}

Amplitude of the derived BMR decreases with increasing Fermi energy as $1/|\varepsilon_{F}|^3$. This is also shown in Fig.2e, where the amplitude $\mathcal{A}_{BMR}$ is plotted for several values of the current density $j$. $\mathcal{A}_{BMR}$  acquires quite large values for small $|\varepsilon_F|$.  However, one should bear in mind that the derived formula is valid for $|\varepsilon_{F}|\gg B, \Gamma_0 $. Moreover, when keeping constant current density, the electric field must increase when $|\varepsilon_F|$ decreases.
It is interesting to compare the calculated BMR with that based on the warping model. The latter was proposed by
Zhang and Vignale, and according to their theory, the amplitude of BMR is proportional to the square of parameter $\lambda$ describing the hexagonal warping strength, and grows linearly with $|\varepsilon_F|$.  BMR amplitudes for the model based on hexagonal warping and that obtained in our approach and based on the interplay of CISP and relaxation processes are given by the formulas:
\begin{eqnarray}
\mathcal{A}_{BMR}^{\rm warp} = \left( 36 \pi \frac{g \mu_{B}}{|e| \hbar^{4}} \right) \frac{\lambda^{2}}{v_{F}^{5}} |\varepsilon_{F}| j b, \\
\mathcal{A}_{BMR}^{\rm CISP} = \left( 30 \pi g \mu_{B}\frac{\hbar^{2}}{|e|} \right) \frac{v_{F}}{|\varepsilon_{F}|^3} j b.
\end{eqnarray}
Rough estimation for Bi$_2$Se$_3$,  assuming  $v_{F} = 5 \cdot 10^{5}$m/s, $\lambda = 165$ eV$\mathrm{\AA}^{3}$, $g = 2$, $b = 5$T and $j = 10$\, A/m, gives  $\mathcal{A}_{BMR}^{\rm warp} = 4.9 \cdot 10^{-6}$ and $\mathcal{A}_{BMR}^{\rm CISP} = 4.40\cdot10^{-5}$ for $\varepsilon_{F} = 0.256$ eV. In turn, for  $\varepsilon_{F} =0.02$ eV we find $\mathcal{A}_{BMR}^{\rm warp} = 3.8 \cdot 10^{-7}$ and $\mathcal{A}_{BMR}^{\rm CISP} = 9.2\cdot10^{-2}$.
%
%
Comparison of both models is shown in Fig.2e, from which follows that while  the contribution to  BMR due to hexagonal warping dominates only at  large values of $|\varepsilon_F|$ (where hexagonal warping is relevant), the contribution based on the model proposed here dominates at smaller values of $|\varepsilon_F|$.

\vspace{0.2cm}
{\textit{Summary and conclusions.}} -- In this letter we have proposed  a new mechanism of the bilinear magnetoresistance due to surface electronic states in topological insulators. This mechanism is based on  the interplay of  non-equilibrium  pseudo-magnetic field due to current and momentum dependent scattering processes. Upon the gauge transformation, which removes the shift of Dirac cone by the in-plane magnetic field, the scattering term depends on both electric and magnetic fields.
We have shown that this leads
to unidirectional behavior of the relaxation time, which affects the current vertex correction.
This is in contrast to the 2D electron gas in semiconductor heterostructures (with Rashba or Dresselhaus spin-orbit interaction), where the mechanism proposed here, and mediated by non-equilibrium spin-polarization, leads to BMR even in the presence of the simplest point-like scalar impurity potential~\cite{ADJBAF_tbp}.

The amplitude of BMR in the mechanism proposed here decreases with increasing  $|\varepsilon_F|$, and thus the mechanism is dominant at relatively  low Fermi energies, contrary to the mechanism based on hexagonal warping, which dominates only at high Fermi energies. In turn,  the $b^2$-magnetoresistance is isotropic within the assumed model.
The theoretical formulation can be  extended further to include
a more accurate description of the band structure (e.g., hexagonal warping and $k^2$ terms) and other scattering potentials.

\vspace{0.2cm}
{\textit{Acknowledgement.}} --
One of us (AD) would like to acknowledge support of German Research Foundation (DFG) through the project SFB 726 and useful discussions with Manuel Bibes.


\clearpage
\onecolumngrid
\section*{}

\appendix
\section{SUPPLEMENTARY MATERIAL}
\subsection{S1. Nonequillibrium spin polarization}
\label{S1}

We consider here the  Edelstein effect in TIs in the absence  of magnetic field, and also analyze  origin of the current-induced effective exchange  Hamiltonian, $\hat{H}_{E}=\mathcal{J} \mathbf{S} \cdot \bsig$.
\subsubsection{Current-induced spin polarization}
In the single-loop approximation the non-equilibrium spin polarization (induced by electric field/current) can be found from the following formula:
\begin{equation}
\label{Sy}
S_{y} = \frac{e \hbar}{2\pi} E_{x} \mathrm{Tr}\int\frac{d^{2} \bk}{(2\pi)^{2}} \hat{S}_{y} \bar{G}^{R}_{\bk 0}(\varepsilon) \hat{v}_{\bk x}\bar{G}^{A}_{\bk 0}(\varepsilon),
\end{equation}
where $\bar{G}^{R/A}_{\bk 0}$ is the impurity-averaged retarded/advanced Green's function of the TI   (for $\mathbf{B} = 0$):
\begin{eqnarray}
\label{GR_TIB0}
G^{R}_{\bk 0} = \frac{\varepsilon \sigma_{0}  + v (k_{x} \sigma_{y} - k_{y} \sigma_{x})}{[ \varepsilon - v k + i \Gamma_{0} ] [ \varepsilon + v k+ i \Gamma_{0}]}.
\end{eqnarray}

Upon substituting Eq.\ref{GR_TIB0} into Eq.\ref{Sy} and making the integration, the  spin polarization takes the form:
\begin{equation}
\label{S_Ed}
S_{y} = \frac{e}{8 \pi} E_{x} \frac{\varepsilon}{v} \tau_{0},
\end{equation}
where $\tau_0 =\hbar /(2\Gamma_0)$.
In turn, the electric current in the TI for zero  magnetic field  is given in single-loop approximation by the following formula:
\begin{equation}
\label{jx}
j_{x} = \frac{e^{2}}{\hbar} \frac{\varepsilon}{4 \pi \hbar}E_{x} \tau_{0}.
\end{equation}

Combining (\ref{jx}) with (\ref{S_Ed}) one finds:
\begin{equation}
\label{Sy-jx}
S_{y} = \frac{\hbar^{2}}{2 e v} j_{x}.
\end{equation}

\subsubsection{An effective Hamiltonian}

The current-induced spin polarization exerts a torque on local electronic spins. This torque may be expressed through an exchange interaction of the induced spin polarization and electron system.  Below we derive the  corresponding effective Hamiltonian.

The simplest form of the Hamiltonian describing surface states of TIs has the form:
\begin{equation}
\hat{H}_{TI} = \pm v (\mathbf{k}\times \hat{\mathbf{z}})\cdot \bsig,
\end{equation}
where the sign $\pm$ depends on spin chirality and the index $\bk$ at the hamiltonian has been dropped for notation simplicity.
Under external electric field the Fermi contour is shifted by $\Delta \mathbf{k} \sim \mathbf{j}$, and
\begin{equation}
\hat{H}_{TI} \rightarrow \hat{H}_{TI\,E} =   \pm v ((\mathbf{k}+\Delta \mathbf{k}) \times \hat{\mathbf{z}})\cdot \bsig = \hat{H}_{TI} + \hat{H}_{E},
\end{equation}
where the electric-dependent term reads
\begin{equation}
\hat{H}_{E} = \pm v (\Delta \mathbf{k} \times \hat{\mathbf{z}})\cdot \bsig = \mp v \Delta{\mathbf{k}} \cdot (\bsig \times \hat{\mathbf{z}}).
\end{equation}
Since $\mathbf{j} = \frac{e}{4\pi} k_{F} v_{F} \Delta \mathbf{k}$ we find $\Delta\mathbf{k} = \frac{8 \pi}{\hbar k_{F}} \mathbf{S}\times\hat{\mathbf{z}}$ and
\begin{equation}
\hat{H}_{E} = \mp \frac{8 \pi v}{\hbar k_{F}} (\mathbf{S}\times\hat{\mathbf{z}} )\cdot (\bsig \times \hat{\mathbf{z}}) = \mp \frac{8 \pi v_{F}}{ k_{F}} \mathbf{S} \cdot \bsig = \mathcal{J} \mathbf{S} \cdot \bsig .
\end{equation}
Thus, the above Hamiltonian appears as a consequence of self-consistent calculations. In the zero-order we assume that $\hat{H}_{E} = 0$ ($\mathbf{S} = 0$) and calculate $\mathbf{S}$. In the 1st order   $\hat{H}_{E} $ is defined by $\mathbf{S}$ given by Eq.(\ref{S_Ed}). Since it is known that the current-induced spin polarization in the presence of exchange field or external magnetic field is only slightly modified relative to CISP in zero fields (see e.g. [A. Dyrdal {\it et al},
Phys. Rev. B \textbf{95}, 245302 (2017)]), in the calculations one can safely assume that $\mathbf{S}$ is defined by Eq.(\ref{S_Ed}).

\subsection{S2. Self-energy and relaxation time}
\label{S2}

Upon the gauge transformation, the total Hamiltonian takes the form
\begin{equation}
\hat{H}^{tot}_{\bq\bq'} = \hat{H}^{0}_{\bq}\delta_{\bq,\bq'}  + \hat{V}_{\mathbf{q} \mathbf{q}'}^{\mathrm{sc}},
\end{equation}
with the unperturbed Hamiltonian $\hat{H}^{0}_{\bq}$ defined as
\begin{equation}
\label{H_0App}
\hat{H}^{0}_{\bq} = v(\mathbf{q} \times \hat{\mathbf{z}})\cdot \bsig = v (q_y \sigma_{x} -  q_x \sigma{y}),
\end{equation}
and the scattering term described by $ \hat{V}_{\mathbf{q} \mathbf{q}'}^{\mathrm{sc}} $.
The Green function corresponding to the Hamiltonian~(\ref{H_0App}) has the form:
\begin{equation}
G_{\mathbf{q}}^{R, 0} = g_{\mathbf{q}0}^{R}  \sigma_{0} + g_{\mathbf{q} x}^{R}  \sigma_{x} + g_{\mathbf{q} y}^{R}  \sigma_{y},
\end{equation}
\vspace{-0.5cm}
\begin{subequations}
\begin{align}
   g_{\mathbf{q}0}^{R}  &=  \frac{1}{2} \left[ G^{R}_{q +} - G^{R}_{q -} \right]\label{g0},\\
   g_{\mathbf{q} x}^{R} &= \frac{q_{y}}{2q} \left[ G^{R}_{q +} - G^{R}_{q -} \right] \label{gx},\\
   g_{\mathbf{q} y}^{R} &= -\frac{q_{x}}{2q} \left[ G^{R}_{q +} - G^{R}_{q -} \right] \label{gy},
\end{align}
\end{subequations}
and $G^{R}_{q \pm} = [\varepsilon - \varepsilon_{\pm} + i \gamma]^{-1}$ with $\varepsilon_{\pm} = \pm v q = \pm v \sqrt{q_{x}^{2} + q_{y}^{2}}$ being the eigenvalues of the Hamiltonian~(\ref{H_0App}), while $\gamma \to 0^{+}$.

The Self-energy in the Born Approximation is defined as:
\begin{eqnarray}
\label{SE-def_App}
\Sigma_{\bq}^{R}(\varepsilon)
=   \int \frac{d^{2} \mathbf{q}'}{(2\pi)^{2}}  \hat{V}^{\mathrm{scatt.}}_{\mathbf{q} \mathbf{q}'} G_{\mathbf{q}'}^{0R} \, \hat{V}^{\mathrm{scatt.}}_{\mathbf{q}' \mathbf{q}} .
\end{eqnarray}

Inserting Eq.(\ref{Halpha_transf}) into Eq.(\ref{SE-def_App}) one gets:
\begin{equation}
\Sigma_{\mathbf{q}}^{R} = \Sigma_{\mathbf{q}0}^{R}\sigma_{0} +  \Sigma_{\mathbf{q} x}^{R}\sigma_{x} +  \Sigma_{\mathbf{q} y}^{R}\sigma_{y} ,
\end{equation}
\begin{eqnarray}
 \Sigma_{\mathbf{q}0}^{R} = - i \gamma_{0} \left[ \left(q^{2} + \frac{\varepsilon^{2}}{v^{2}} \right) + \frac{B^{2}}{v^{2}} + \frac{2}{v^{2}} B_{y} \mathcal{J} S_{y} - \left( \frac{q_{y}}{v} B_{x} - \frac{q_{x}}{v} (B_{y} + \mathcal{J} S_{y})\right) \right] ,\\
\Sigma_{\mathbf{q} x}^{R} = \mp i \gamma_{0} \frac{|\varepsilon|}{2 v} \left(q_{y} - 2 \frac{B_{x}}{v} \right) , \hspace{7.0cm}\\
\Sigma_{\mathbf{q} y}^{R} = \pm i \gamma_{0} \frac{|\varepsilon|}{2 v} \left( q_{x} + 2 \frac{B_{y} + \mathcal{J} S_{y}}{v}\right) ,\hspace{6.1cm}\\
\end{eqnarray}
where  $\gamma_{0} = \frac{n_{i} \alpha^{2} |\varepsilon|}{4 v^{2}}$, and the sign in front of $\Sigma_{\mathbf{q} x/y}^{R}$ corresponds to the positive/negative $\varepsilon$ respectively.
In the basis of eigenstates, the corresponding selfenergy reads (note, the selfenergy is imaginary)
\begin{equation}
\label{SigPl}
\Sigma_{\mathbf{q} \pm}^{R} \equiv - i \Gamma_{\pm} =  \Sigma_{\mathbf{q}0}^{R}  \pm \left( \frac{q_{y}}{q} \Sigma_{\mathbf{q}x}^{R} - \frac{q_{x}}{q} \Sigma_{\mathbf{q} y}^{R}\right) .
\end{equation}
Note, that $\Gamma_{+}(\varepsilon )= \Gamma_{-}(-\varepsilon )$.
Accordingly, the relaxation rate for  quasiparicles in the conduction band ($\varepsilon > 0$) is the same as the relaxation rate for quasiparicles in the valence band ($\varepsilon <0$), i.e. $\Gamma_{+}(\varepsilon ) = \Gamma_{-}(-\varepsilon ) = \Gamma$, and $\Gamma$ takes the following form:
\begin{equation}
\Gamma(q, \phi) = \frac{\gamma_{0}}{4 v^{2}} \left[ (|\varepsilon| + v q)^{2} + 8 B_{y} \mathcal{J} S_{y} + 4 B^{2} + 4 (|\varepsilon |+ q v) \left( (B_{y} + \mathcal{J} S_{y}) \cos(\phi) - B_{x} \sin(\phi)\right)\right].
\end{equation}
Taking this expression at the mass surface, $v q = |\varepsilon|$ we get:
\begin{eqnarray}
\Gamma = \Gamma_{0} \left[ 1 + 2 \frac{B_{y} \mathcal{J} S_{y}}{\varepsilon^{2}} + \frac{B^{2}}{\varepsilon^{2}}
+ \frac{2}{|\varepsilon|} \left[ (B_{y} + \mathcal{J} S_{y}) \cos\phi - B_{x} \sin\phi\right]\right] ,
\end{eqnarray}
where $\Gamma_{0} = n_{i} \lambda^{2} \frac{|\varepsilon|^{3}}{4 v^{4}}$, i.e Eq.(10).

\subsection{S3. Conductivity and vertex correction}
\label{S3}

\subsubsection{Vertex correction}
The impurities vertex correction for the velocity will be found self-consistently based on the vertex equation:
\begin{equation}
\label{VertexEq}
\mathcal{V}_{\bq x} = \hat{v}_{\bq x} + \int \frac{d \bq'^{2} }{(2\pi)^{2}} \hat{V}^{\alpha}_{\bq \bq'}\bar{G}_{\bq'}^{R} \mathcal{V}_{\bq'x} \bar{G}_{\bq'}^{A}\hat{V}^{\alpha}_{\bq' \bq} .
\end{equation}
Since  $\hat{v}_{\bq x} = - \frac{v}{\hbar} \sigma_{y}$, one can assume that the renormalized velocity vertex function $\mathcal{V}_{\bq x}$ takes the form:
\begin{equation}
\label{VxPost}
\mathcal{V}_{\bq x} =a q_{x} \sigma_{0} + b \sigma_{x} + c \sigma_{y} + d \sigma_{z}.
\end{equation}
Inserting (\ref{VxPost}) into (\ref{VertexEq}) we get:
\begin{equation}
 a \sigma_{0} + b \sigma_{x} + c \sigma_{y} + d \sigma_{z} = \sigma_{y} + \sum_{l = 0, x, y, z} \int \frac{d {\bq'}^{2}}{(2\pi)^{2}} \mathcal{W}_{l} \sigma_{l} \frac{\pi}{4 \varepsilon^{2}}  \left[ \frac{1}{\Gamma_{\bq' +}} \delta(\varepsilon - v q') + \frac{1}{\Gamma_{\bq' -}} \delta(\varepsilon + v q') \right] ,
\end{equation}
where $\mathcal{W}_{l} = [\hat{V}^{\alpha}_{\bq \bq'}g_{\bq'}^{R} \mathcal{V}_{\bq' x} g_{\bq'}^{A}\hat{V}^{\alpha}_{\bq' \bq}]_{l}$. The above equation leads to the set of four algebraic equations for coefficients $a, b, c, d$:
\begin{subequations}
	\begin{equation}
	2 a = a F_{0a} + b F_{0b} + c F_{0c} + d F_{0d},
	\end{equation}
	\begin{equation}
	2 b = a F_{xa} + b F_{xb} + c F_{xc} + d F_{xd},
	\end{equation}
	\begin{equation}
	\hspace{0.8cm}2 c =  a F_{ya} + b F_{yb} + c F_{yc} + d F_{yd} - 2\frac{v}{\hbar},
	\end{equation}
	\begin{equation}
	2 d = a F_{za} + b F_{zb} + c F_{zc} + d F_{zd}.
	\end{equation}
\end{subequations}
Here we use the following notation:
\begin{eqnarray}
F_{m l} = \int \frac{dq' d\phi'}{(2\pi)^{2}} \frac{\pi q'}{4 \varepsilon^{2}} {\mathrm{Tr}} \left\{ \sigma_{m} \mathcal{W}_{l} \sigma_{l} \right\}  \left[ \frac{1}{\Gamma_{\bq' +}} \delta(\varepsilon - v q') + \frac{1}{\Gamma_{\bq' -}} \delta(\varepsilon + v q') \right]
\end{eqnarray}

Next we focuse on the case of $\varepsilon > 0$, thus
\begin{eqnarray}
F_{m l} = \int \frac{dq' d\phi'}{(2\pi)^{2}} \frac{\pi q'}{4 \varepsilon^{2} \Gamma_{\bq' +}} {\mathrm{Tr}} \left\{ \sigma_{m} \mathcal{W}_{l} \sigma_{l} \right\}   \delta(\varepsilon - v q')
\end{eqnarray}
The solution of the set of equations is:
\begin{eqnarray}
d = 0,\hspace{12.1cm}\\ \label{a}
a = \frac{v}{4 q \hbar |\varepsilon|^{3}}  \left[ - qv \left( B_{x}^{2} + 5 By(By + \mathcal{J} S_{y}) - \varepsilon^{2} \right) \right] + v \frac{B_{y} + \mathcal{J} S_{y}}{4 q_{x} \hbar |\varepsilon|^{3}} \left( \varepsilon^{2} - v^{2} q^{2} + 4 B_{x} v q_{y} \right),\\ \label{b}
b = - \frac{ v^{2}}{4 \hbar \varepsilon^{4}} q_{x} \left[ 2 B_{x} \varepsilon^{2} + v q_{y} (B^{2} + 2 B_{y} \mathcal{J} S_{y} - \varepsilon^{2})\right],\hspace{5.5cm} \\ \label{c}
c =  \frac{v}{8 \hbar \varepsilon^{4}} \left[ 5 \varepsilon^{2} (B^{2} + 2 B_{y} \mathcal{J} S_{y}) - \varepsilon^{4} + v^{2}(q_{x}^{2} - q_{y}^{2}) (B^{2} + 2 B_{y} \mathcal{J} S_{y} - \varepsilon^{2}) - 4 \varepsilon^{2} v q_{y} B_{x} \right].\hspace{0.5cm}
\end{eqnarray}

Thus, one can write the renormalized  velocity vertex function as:
\begin{eqnarray}
\label{VxPost}
\mathcal{V}_{\bq x} = a q_{x} \sigma_{0} + b \sigma_{x} + c \sigma_{y} \hspace{0.7cm}\nonumber\\ = v_{x0} \sigma_{0} + v_{xx} \sigma_{x} + v_{xy} \sigma_{y}
\end{eqnarray}
Note that at the mass surface, $v q = |\varepsilon|$ we get:
\begin{eqnarray}
v_{x0}|_{q = \frac{|\varepsilon|}{v}} = \frac{v}{4 \hbar} \left[ 1 - \frac{B^{2}}{\varepsilon^{2}} - 5 \frac{B_{y}\mathcal{J} S_{y}}{\varepsilon^{2}}\right]\, \cos\phi\, + \frac{v}{\hbar} \frac{B_{y}(B_{y}+ \mathcal{J} S_{y})}{\varepsilon^{2}} \sin\phi , \hspace{2.75cm}\\
v_{xx}|_{q = \frac{|\varepsilon|}{v}}  =  \frac{v}{8 \hbar} \left[ 1 - 2 \frac{B_{y} \mathcal{J} S_{y}}{\varepsilon^{2}} - \frac{B^{2}}{\varepsilon^{2}}\right] \sin(2\phi) -
\frac{v}{2 \hbar} \frac{B_{x}}{|\varepsilon|} \cos\phi , \hspace{4.1cm}\\
v_{xy}|_{q = \frac{|\varepsilon|}{v}}  = - \frac{v}{8 \hbar} \left[ 1 - 10 \frac{B_{y} \mathcal{J} S_{y}}{\varepsilon^{2}} - 5 \frac{B^{2}}{\varepsilon^{2}} + \left( 1 - 2 \frac{B_{y} \mathcal{J} S_{y}}{\varepsilon^{2}} - \frac{B^{2}}{\varepsilon^{2}} \right) \cos (2\phi)  + 4 \frac{B_{x}}{|\varepsilon|} \sin\phi \right].
\end{eqnarray}
Equations~(56)-(59) are then used to calculate the diagonal conductivity from Eq.~(\ref{sigma_xx}).

\end{document}